\documentstyle[epsfig]{aipproc}

\begin{document}

\vspace*{-1.7cm}

\begin{flushright}
MC-TH-97/8
\end{flushright}

\vspace*{-1.5cm}

\title{Photoproduction\thanks{Talk presented at the 5th International
Workshop on Deep Inelastic Scattering and QCD, Chicago, April 1997.}}
\author{Jeffrey R. Forshaw}
\address{Department of Physics and Astronomy, University of Manchester,
Manchester. M13 9PL. United Kingdom}

\maketitle

\vspace*{-0.5cm}

\begin{abstract}
Some selected topics in photoproduction are reviewed. In particular, the
focus is on quasi-real photon--proton scattering as measured at
HERA. Single \& dijet rates and jet profiles are discussed, as are
open charm production and inelastic charmonium production.
The many intriguing theoretical issues which can be addressed through
study of these processes are highlighted.
\end{abstract}

\vspace*{-0.3cm}

\section*{Introduction}
In this talk, I'm going to present a summary of recent developments in
photoproduction at HERA. By ``photoproduction'' I mean $\gamma p$
interactions in which the photon is close to being on its mass shell.
Furthermore, I shall focus on processes in which there exists at least
one hard scale -- so that we have the chance to use perturbation theory.
I'm going to divide the talk into 4 parts: jets (single \& dijet rates);
jet profiles; open charm production ($D^*$'s); inelastic charmonium
production. I won't talk about processes with rapidity gaps, since these
diffractive processes are covered in other people's presentations.

\section{Jet cross-sections}
I wish to start by recalling some basic terminology. It is convenient to
talk about two classes of photoproduction event: those in which all of
the energy of the incoming photon is delivered to the hard subprocess 
(these are called ``direct'' processes) and those in which only part of
the incoming photon energy goes into producing the hard subprocess
(these are called ``resolved'' processes). 

\begin{figure} 
\centerline{\epsfig{file=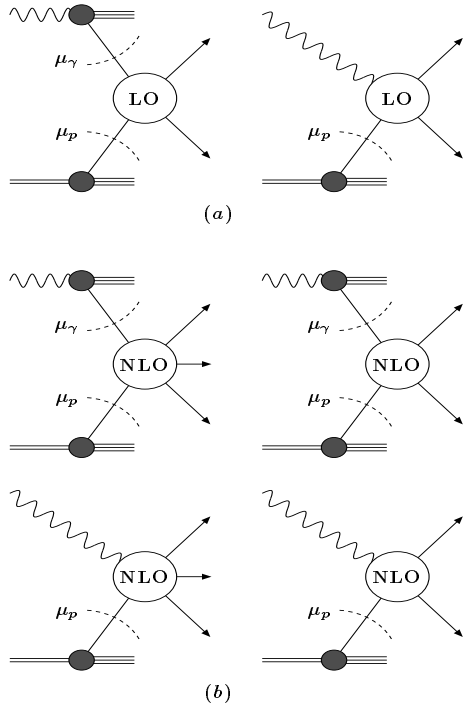,height=3.7in,width=2.5in,bbllx=193pt,%
bblly=292pt,bburx=420pt,bbury=717pt}}
\vspace{10pt}
\caption{LO (a) and NLO (b) direct and resolved contributions to jet
photoproduction}
\label{fig1}
\end{figure}

Fig.\ref{fig1}(a) shows the
lowest order contribution from these two types of process in the case of jet
production. The dotted lines show where we chose to separate the hard
process from the non-perturbative parton distribution functions. The
factorization scales $\mu_{\gamma}$ and $\mu_p$ are in principle
arbitrary but a good choice would be to pick them to be $\sim p_T$
(i.e. the jet transverse momentum) since this choice will account for
the fact that the incoming partons have a large region of transverse
phase space into which they can radiate other partons. Nonetheless, it
is certainly true that the theoretical calculation will exhibit a strong
sensitivity to any variation in the factorization scales since 
the $2 \to 2$ hard subprocesses contain no dependence upon the
factorization scales whereas the parton distribution functions do.
For example,
we could effectively turn off all parton evolution by picking the
factorization scales to be much smaller than $p_T$. At this level of
calculation, it is clear what we mean by ``direct'' and ``resolved''
processes. 

In Fig.\ref{fig1}(b), the ${\cal O}(\alpha_s)$ corrections to the jet
production process are shown. We have to account for the fact that the
hard subprocess can radiate off 3 partons into the final state as well
as the virtual corrections to the 2 parton final state. Again, we can
classify diagrams as being direct or resolved. However, there are
collinear divergences associated with the 3 parton final state. This
means that the 3 parton hard subprocesses are dependent upon the
factorization scales. Let us focus on the $\mu_{\gamma}$ dependence.
By varying $\mu_{\gamma}$ we vary the amount of the next-to-leading
order (NLO) $2 \to 3$ processes that should really be attributed to the 
lowest order (LO) resolved process. In particular, the amount of NLO
direct that should rightfully be factorized into the LO resolved
contribution depends upon the choice of $\mu_{\gamma}$. In this way, the
strong dependence upon the factorization scales is reduced by including
the NLO contribution. Also, we can no longer uniquely define the
separation of direct and resolved processes. 

\subsection{Single jet inclusive}

\begin{figure}[t] 
\centerline{\epsfig{file=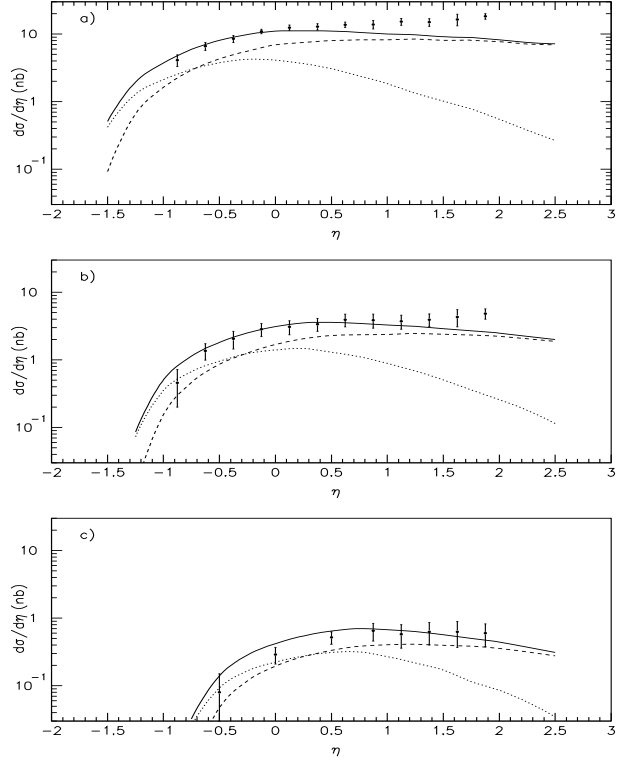,height=4.0in,width=3.2in}}
\vspace{10pt}
\caption{Inclusive single jet cross-section integrated above $E_T=8$ GeV
(a), 11 GeV (b) and 17 GeV (c). The data \protect\cite{1jetZeus} are compared
with the NLO calculation of \protect\cite{KKS}. Dotted line shows the direct
contribution, dashed line the resolved and solid the total.}
\label{fig2}
\end{figure}

NLO calculations of the single jet inclusive rate have been available
for a while now \cite{jets}. 
Comparison with the data \cite{1jetH1,mievi,1jetZeus} should tell us about the
parton content of the photon, and in particular its gluonic content
(which is not really constrained from elsewhere). However, in just the
region where one expects sensitivity to the gluon in the photon the
NLO theory is not working! Forward jets (i.e. those heading largely in the
proton direction and hence having positive pseudo-rapidity, $\eta$) with
modest values of transverse energy, $E_T$, are most sensitive to the
gluon content of the photon. Fig.\ref{fig2} compares the 1993 ZEUS data
with the theoretical calculations of Klasen, Kramer \& Salesch
\cite{KKS} which use the GS(HO) photon parton densities \cite{GSpdf}
and the MRSD- proton densities \cite{MRS}. Comparison
reveals that the theory is well below the data for
forward jets at the lower $E_T$ values (this conclusion is also
supported by the H1 data and the ZEUS 1994 data). 
The problem seems not so bad as $E_T$ is
increased. One might argue that the difference between theory and data
should be attributed to the fact that the photon parton density
functions need adjusting, i.e. more gluons are required at low
$x$. However, one should be cautious. These forward jets are also much
broader than expected from Monte Carlo studies. 

It is possible that the broadening of the forward jets comes about
in the presence of a large underlying event. Models which attempt to
account for some or all of this physics \cite{mis,BFS} 
have received some support from the data \cite{mievi}. 
In particular, simple models of multiple parton
scattering produce large effects in the forward region, e.g. see
\cite{BFS}. Multiple parton scattering
is anticipated on the grounds that forward jets at low $E_T$ are
produced as a result of interactions between slow partons in the
colliding particles. We know that QCD predicts a proliferation of these
slow partons, and as such it may well be that more than one pair of
them can interact in each $\gamma p$ interaction.

In summary, the HERA experiments are providing precise data which is
ready for comparison with the theory. However, before we can safely
extract the photon parton distribution functions we really need to
understand what is going on with the forward jets. I think the issue of
the forward jets is potentially very interesting. If multiple
interactions really are present, then we are looking at a new regime of
QCD. 

\subsection{Dijets}

Data on two or more jets \cite{2jetZeus,angle,2jetH1} provides us with further 
options to test QCD and
understand the nature of the ``strongly interacting'' photon. In
particular, the two highest $E_T$ jets can be used to compute the
variable 
\begin{equation}
x_{\gamma} = \frac{ \sum_{{\rm jets}} E_T e^{-\eta}}{2 y E_e}.
\end{equation}
At lowest order, $x_{\gamma}=1$ for direct events and $x_{\gamma} < 1$
for resolved events. A cut on $x_{\gamma}$ therefore provides a physical
definition of direct and resolved event classes. ZEUS has defined 
direct enriched and resolved enriched samples by separating events
according to a cut at $x_{\gamma} = 0.75$. The direct enriched sample is
very sensitive to the small-$x$ gluon content of the proton: the more
backward the dijets, the lower the $x$ values in the proton that are
probed. Conversely, the resolved enriched sample is sensitive to the
gluon content in the photon. In addition, NLO calculations for the
dijet rates are now available for comparison with the data \cite{KK,O}.
Let's summarize the situation as it stands right now.

\begin{figure}[t] 
\centerline{\epsfig{file=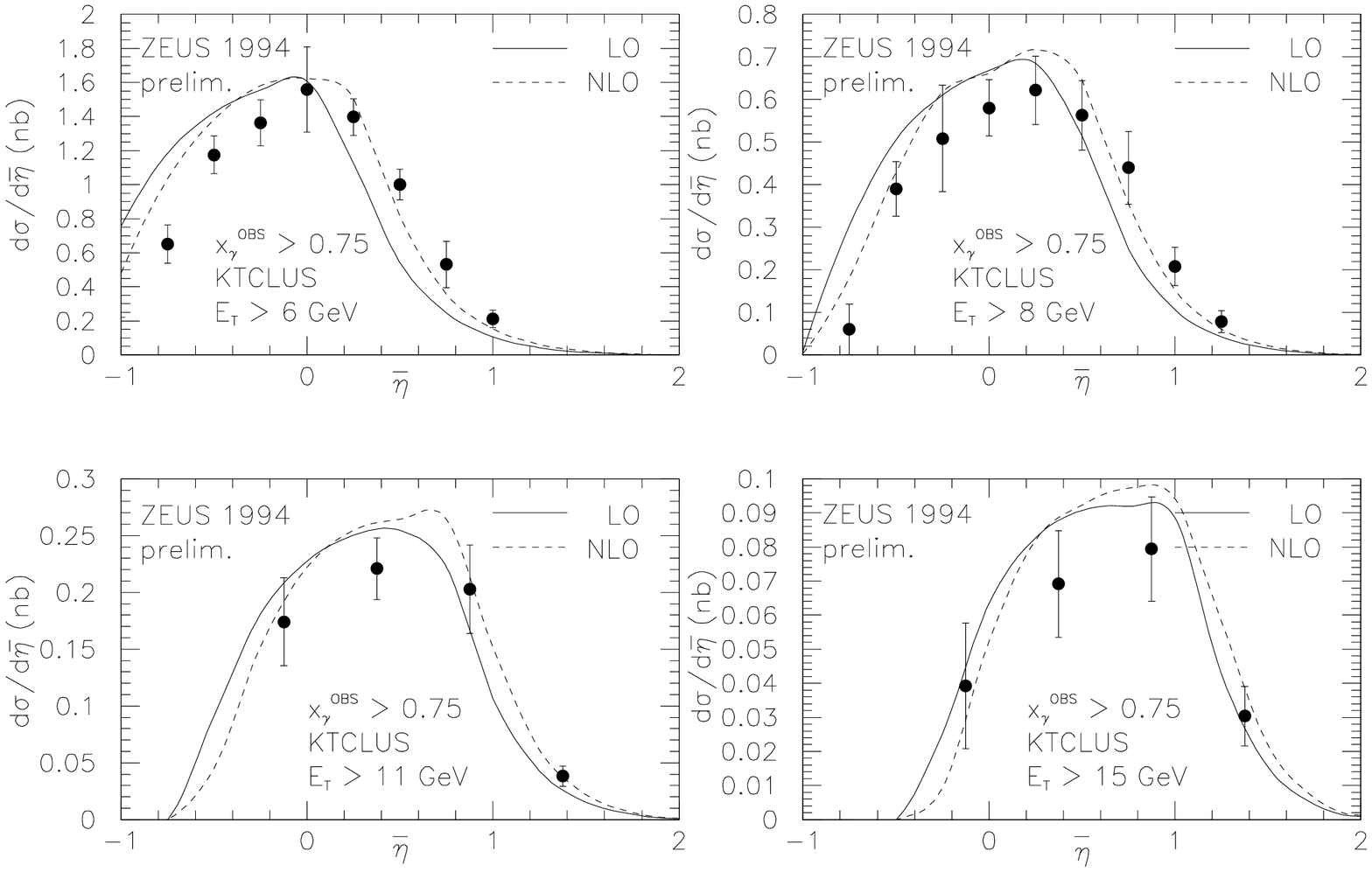,height=3.6in,width=2.0in,angle=270}}
\centerline{\epsfig{file=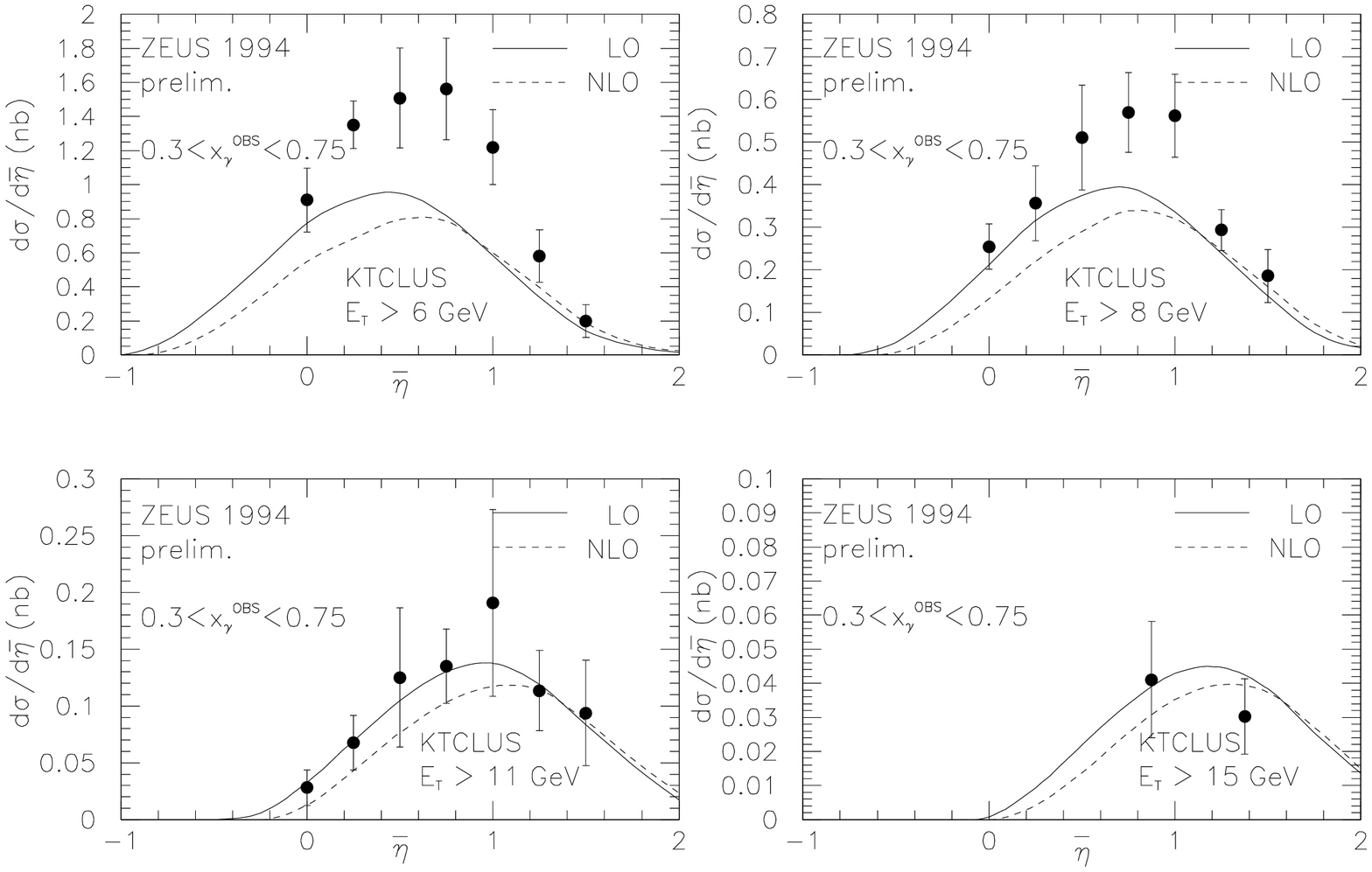,height=3.6in,width=2.0in,angle=270}}
\vspace{10pt}
\caption{Inclusive dijet cross-section for direct and resolved enriched
processes as a function of the average jet rapidity, $\bar{\eta}$, and
integrated over the rapidity difference, $\Delta \eta$, of the two
highest $E_T$ jets, i.e. $|\Delta \eta| < 0.5$, and $E_T$ as shown. The
LO and NLO theory of \protect\cite{KK} is compared to the data
\protect\cite{2jetZeus}. 
The GRV \protect\cite{GRV} and CTEQ3M \protect\cite{CTEQ} 
NLO ${\overline{MS}}$ parton densities were used.} 
\label{fig3}
\end{figure}

For $x_{\gamma} > 0.75$, as can be seen from Fig.\ref{fig3}, the NLO
theory does a good job. However (and this is not shown in the plot) 
there remains quite a large contamination from the large-$x$
part of the photon quark distribution functions. This arises because of
the harder form of the photon quark densities. To unravel the effects of
the low-$x$ gluons in the proton from the large-$x$ quarks in the photon
requires a tighter cut on $x_{\gamma}$. This is feasible, for example
for $E_T > 30$ GeV and 250pb$^{-1}$ of $ep$ data one expects around 4500
events with $x_{\gamma} > 0.9$ \cite{HERAwkp}. Also, comparison between
data and theory requires that the same jet algorithm is used. To
facilitate a clean comparison between data and theory, the ZEUS
collaboration has started to used the $k_t$-cluster algorithm \cite{ktclus}.

For $x_{\gamma} < 0.75$ Fig.\ref{fig3} reveals that the theory falls
well below the data for the lowest $E_T$ forward dijets. The
effect exhibits a strong dependence upon the $E_T$ cut, which suggests
that it cannot be explained by modifying the parton distribution functions of
the photon in any sensible way. Presumably, this is the same problem
as that which we encounter for the single jets. Again, multiple
interactions help fix the problem.  

The dijet measurements shown in Fig.\ref{fig3} 
have been made with a cut on the
minimum $E_T$ of the jets and the cut is the same for both jets. This
introduces a further theoretical problem. This arises because most of
the jets will be produced around the minimum allowable $E_T$, i.e. the
typical difference between the jet transverse momenta, $\Delta p_T$, will
be small. So, the 3
parton final state (which is present in the NLO calculation) must have
one of the partons either collinear with another, or very soft. The
collinear configuration is easy to deal with (it is factorized) but the
soft parton emission leads to a $\ln \Delta p_T$ contribution. This
large logarithm signals that multiple soft parton emission is
important. These soft parton effects can be
studied by looking explicitly at the $\Delta p_T$ distribution of the
dijets or they can be avoided by making a cut which keeps away from
$\Delta p_T \approx 0$.

\begin{figure} 
\centerline{\epsfig{file=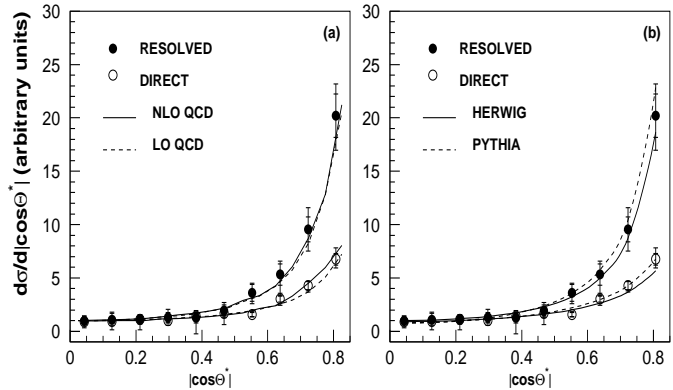,height=2.5in,width=3.5in}}
\vspace{10pt}
\caption{Dijet angular distribution}
\label{fig4}
\end{figure}

Complementary to the jet rapidity and $E_T$ distributions is the
distribution in the polar angle of the jets, as defined in the dijet
centre-of-mass frame. Such a measurement is insensitive to the parton
density functions but sensitive to the subprocess which drives the jet
production. Fig.\ref{fig4} 
shows the data \cite{angle} and the good agreement with NLO theory \cite{O}. 
The steepening of the $\cos \Theta^*$ distribution as $|\cos
\Theta^*| \to 1$ is greater in the resolved enriched sample than in the
direct enriched sample due simply to the fact that the resolved process
proceeds predominantly through $t$-channel gluon exchange whilst the
direct processes go via $t$-channel quark exchange.

In conclusion, the dijets provide information which complements the
single jet measurements. The data is now reaching a high level of
precision, and comparison with NLO theory has revealed a number of pressing
issues. In particular, we need to understand better the forward jets,
ensure that we are using the most convenient jet algorithm, and be aware
of any sensitivity from soft parton emission. Once these issues have been
addressed, we can expect to gain further insight into the gluon content
of both the photon and proton.  

\section{Jet Shapes}
Now I want to address the question: ``What is a jet?''. An observable
which has been used to investigate this
question is the shape variable $\psi(r,R,E_T,\eta)$ which is the
fraction of the jet (of cone size $R$) energy lying within a cone of
size $r$ \cite{EKS}. 
The lowest non-trivial order for this quantity involves the 3
parton matrix elements. To compute to order $\alpha_s^2$ requires the 4
parton final state matrix elements \cite{GK}. 
Klasen \& Kramer have been able to fit the ZEUS data
on jet shapes \cite{jetshapes} but only after introducing the parameter,
$R_{{\rm sep}}$, into their calculations \cite{KlK}. 
Let's take a closer look at the $R_{{\rm sep}}$ parameter. 

\begin{figure} 
\centerline{\epsfig{file=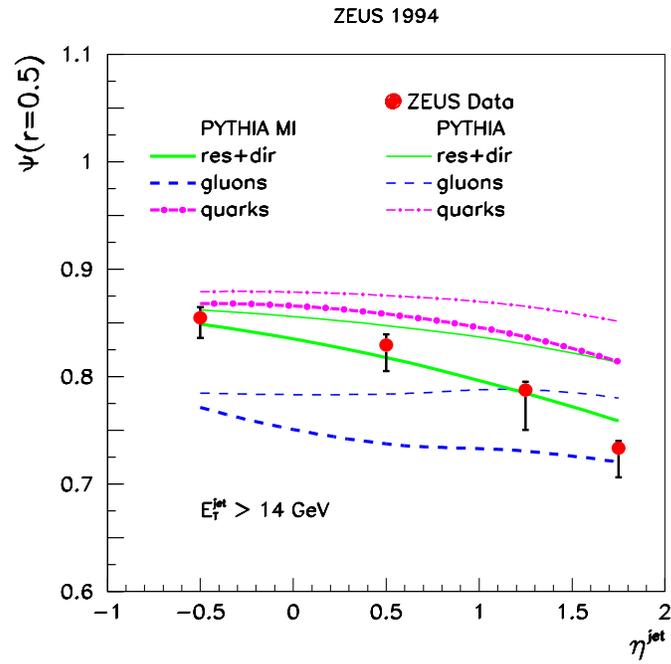,height=3.5in,width=3.5in}}
\vspace{10pt}
\caption{Jet shape from single jet production. Comparison between data
\protect\cite{jetshapes} and PYTHIA, both with and without multiple parton scattering
(MI), is shown.}
\label{fig5}
\end{figure}

It's a good idea to first remind ourselves of the steps involved 
in building jets using a cone algorithm \cite{cone}.  The ZEUS jets
are found using an iterative cone algorithm (PUCELL). 
This algorithm defines some threshold energy, $E_0$, and 
calorimeter cells above this energy define the seeds for the jet finding.
A jet is formed by summing all cells within a distance $R$ (in
$\eta-\phi$ space) of the seed cell. If, after this, the jet axis no
longer coincides with that of the original seed cell then the new jet
axis is used and all cells within $R$ of it are used to redefine the
jet. This process is iterated until a stable jet direction is
achieved. It is possible that some of the jets overlap and so some
criterion needs to be established to decide what should be done with
these jets. For example, if one of the overlapping jets has more than
75\% of its energy in common with another jet then one could merge the
two jets, whilst if the overlap energy is less then one could split the two
jets by assigning particles to the jet to which they are
nearest. Finally, jets are accepted if their $E_T$ is larger than some
minimum value. In the theoretical case, two partons are combined if they
lie within a distance $R_{ij}$ of each other, where
\begin{equation}
R_{ij} \leq {\rm min} \left[
\frac{(E_{Ti}+E_{Tj})R}{{\rm max}(E_{Ti},E_{Tj})},R_{{\rm sep}}\right]. 
\end{equation}
We have introduced $R_{{\rm sep}}$ into this definition, it is introduced
so as to simulate the effect of seed finding. For example, two partons
with equal $E_T$'s lying a distance $2R$ apart could be considered to
form a single jet (the partons lying on the edge of the jet cone). Such
a configuration could not occur in the experimental algorithm, since
there is no seed cell in between the two partons. As a result, for a
3 (or less) parton final state, picking  $R_{{\rm sep}} = R$ should
match the theoretical and experimental definition of the jets. Notice
that the 3 parton final state does not call for any jet merging. By
fitting $R_{{\rm sep}}$ to the data, one is accounting for higher
order effects by fitting a single parameter (to fit the data Klasen \&
Kramer need to vary $R_{{\rm sep}}$ from 1.3 to 1.8 for jets with
$\eta \approx -1$ to $\eta \approx 2$ respectively). For example, parton shower
and hadronization effects (plus, maybe, multiple parton scattering) all
effect the final state and we do not really learn much about them by
tuning the theoretical calculation.  Fig.\ref{fig5} demonstrates that forward
jets are broader, as I claimed earlier. It also shows how multiple
parton interactions can help improve the situation. It really is a
challenge to theorists to explain the variation of $R_{{\rm sep}}$.

\begin{figure} 
\centerline{\epsfig{file=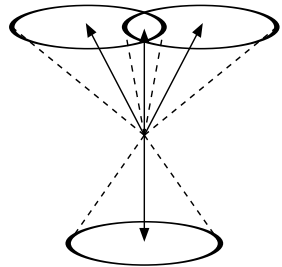,height=1.5in,width=1.5in,%
bbllx=221pt,bblly=452pt,bburx=323pt,bbury=545pt}}
\vspace{10pt}
\caption{A four parton final state}
\label{fig6}
\end{figure}

Steps in this direction require a truly NLO calculation of the jet shape
variable. As mentioned above, this means we need to consider the 4
parton final state. At this point, we realise that the iterative cone
algorithm is not infra-red safe. By this we mean that it is not insensitive
to the addition of a soft parton \cite{GK}. To see this consider the 4
parton final state illustrated in Fig.\ref{fig6}. 
The middle of the upper 3 arrows denotes a soft
parton, which has energy just above the threshold energy, $E_0$. The
presence of this seed in between the two hard partons would cause all 3
partons to be merged into a single jet. So our shape variable is
sensitive to the cell threshold energy. This is an acute problem at any
fixed order in perturbation theory (beyond LO). However, a strong $E_0$
dependence isn't seen in Monte Carlo studies. Let me outline the basic
idea as to why this should be so \cite{GK,MikeS}. 
Denote the probability for emitting a
single parton into the overlap region, $P_1$, where

\begin{equation}
P_1 \sim \alpha_s \ln \frac{E_T}{E_0}.
\end{equation}
This probability exponentiates on summing over all possible numbers of
soft partons, i.e.
\begin{equation}
P_{{\rm all}} \sim 1 - {\rm e}^{-P_1}.
\end{equation} 
So as $E_0 \to 0$ a soft parton is guaranteed to be in the overlap
region, i.e. $P_{{\rm all}} \to 1$. Thus we see that, on expanding the
exponential, any finite order in $\alpha_s$ is sensitive to large $\ln
E_0$ terms but that this sensitivity goes away on summing to all orders
(as is done in the parton shower routines). As pointed out by Steve
Ellis \cite{SE} this problem can be fixed simply be redefining the jet
algorithm, so that the mid-point between jets is always treated as a
seed, regardless of any activity. Alternatively, the $k_t$ cluster
algorithm avoids such problems. I think it is fair to say that studies
on jet structure are reaching the point where theory really is starting to
address the important issues. We should learn much over the coming
few years. 

\section{Open Charm}
The photoproduction of charm quarks at large $p_T$ is a process which
involves two large scales, $p_T, m_c \gg \Lambda$ and as such, makes
life more complicated from the theoretical point of view. Good data,
which can be expected in the future (especially if the charm can be
tagged using a microvertex detector), will surely shed light on this
intriguing area. At present, there are two main routes used in
theoretical calculations. I'll start by describing briefly each.

{\bf Massive Charm:}
The charm quark mass is considered to provide the hard scale, as such
charm only ever appears in the hard subprocess and there is no notion of
radiatively generated charm in the sense of parton evolution
\cite{massive}. This means
that terms $\sim \alpha_s \ln(p_T^2/m_c^2)$ are not summed to all orders
(in the parton distribution and fragmentation functions). As such, we
might expect this approach to become less accurate when $p_T^2 \gg
m_c^2$. However, it does provide a systematic way of accounting for
charm quark mass effects, which will be important for $m_c \sim p_T$.

{\bf Massless charm:}
In this approach, the charm quark is treated as massless (above
threshold), and as such is treated like any other light quark in the
parton evolution equations and hard subprocess cross-sections. 
The $\ln(p_T/m_c)$ terms are now summed to all orders, but charm quark
mass effects are ignored. So, this approach should get better as
$p_T/m_c$ increases.

At present the data \cite{cdatH1,cdatZeus,low} 
lies in the intermediate region where $p_T < 10$
GeV, i.e. it is not clear which, if any, of the two approaches should be
used. In order to compute the inclusive $D^*$ rate, one needs the
appropriate fragmentation function. Either the Peterson \cite{Peterson}
form or the $x^a (1-x)^b$ form do a good job, and have been well
constrained by $e^+ e^-$ data.

\begin{figure}[t] 
\centerline{\epsfig{file=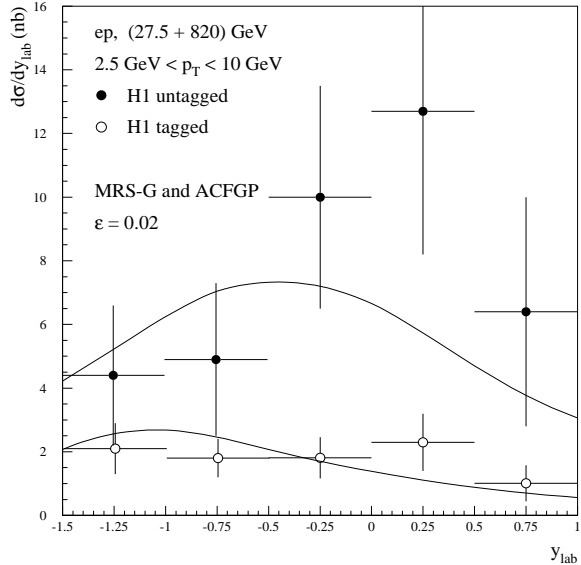,height=3.0in,width=3.0in}}
\vspace{10pt}
\caption{Comparison of the $D^*$ cross-section data \protect\cite{cdatH1} 
with the calculation of \protect\cite{CG}.}
\label{fig7}
\end{figure}

Initial comparison between theory and data suggested that the massive
charm calculation was too low, e.g see \cite{low}. 
More recently, the situation has changed
somewhat \cite{CG,BKKS}. The full NLO calculations require that the
fragmentation function be consistently extracted from the $e^+ e^-$
data. When this is done, Cacciari and Greco find that the theoretical
predictions are increased significantly relative to what is found using
the softer (LO) fragmentation functions \cite{CG}. 
This is true for both massless
and massive charm calculations and, within theoretical uncertainties,
both are now consistent with the present data. For example,
Fig.\ref{fig7} is a plot from \cite{CG}, which shows that the NLO theory
assuming massless charm agrees well with the data.  

More data at large $p_T$ and increased statistics at intermediate $p_T$
will certainly help in our study of the interplay of $\ln(p_T^2/m_c^2)$
and $m_c^2/p_T^2$ effects. In addition, for $p_T \gg m_c$ we have the
possibility to study the ``intrinsic'' charm within the photon   
(charm in dijets offers good prospects here).

\section{Charmonium}
Originally, inelastic photoproduction of charmonium,
e.g. $J/\psi$, was advertised as an ideal way to extract the gluon density
in the proton (since it is driven by photon-gluon fusion into a $Q
\bar{Q}$ pair). More recently, NLO calculations have put a dampener on this
goal (as we shall see shortly). However, there has been a great deal of
recent interest in the non-relativistic QCD (NRQCD) approach to
heavy quarkonium production, and the inelastic photoproduction of heavy
quarkonia provides the ideal opportunity to test NRQCD. 

\begin{figure} 
\centerline{\epsfig{file=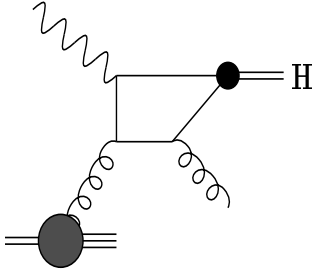,height=1.5in,width=1.7in,%
bbllx=169pt,bblly=388pt,bburx=286pt,bbury=475pt}}
\vspace{10pt}
\caption{Leading contribution to quarkonium photoproduction.}
\label{fig8}
\end{figure}

Bodwin, Braaten and Lepage derived a factorization formula which
describes the inclusive production (and decay) of a heavy quarkonium state
\cite{BBL}. In the case of photoproduction, Fig.\ref{fig8} shows the
lowest order contribution. 
The NRQCD factorization formula for the corresponding
cross-section reads
\begin{equation}
d\sigma(H+X) = \sum_{n} d\hat{\sigma}(Q \bar{Q}[n]+X) \langle O^H_n
\rangle. 
\end{equation}
$X$ denotes that the process is inclusive, $d\hat{\sigma}(Q
\bar{Q}[n]+X)$ is the perturbatively calculable cross-section for
$\gamma p \to Q \bar{Q} + X$ and it can be written as a series expansion
in $\alpha_s(m_Q)$. The $Q \bar{Q}$ pair is produced with
quantum numbers $n$. The matrix element, $\langle O^H_n \rangle$,
contains the long distance physics associated with the formation of the
quarkonium state $H$ from the $Q \bar{Q}$ state -- it is
essentially the probability that the pointlike $Q \bar{Q}$ pair forms
$H$ inclusively. The typical scale
associated with this part of the process is $\sim m_Q v$ which is much
smaller than $m_Q$ ($v$ is the relative velocity of the $Q \bar{Q}$
pair, and is small for heavy enough quarks). This hierarchy of scales
underlies the NRQCD factorization. Note that the $Q \bar{Q}$ state is
not restricted to having the same quantum numbers as the
meson. Fortunately, there exist ``velocity scaling rules'' which allow us
to identify which states, $n$, are the most important. More precisely,
the ``velocity scaling rules'' order the operators $\langle O^H_n
\rangle$ according to how many powers of $v$ they contain,
i.e. relativistic corrections can be computed systematically. 

The NRQCD approach therefore provides us with a systematic way of
computing inclusive heavy quarkonium production (modulo corrections
which are suppressed by powers of $\sim \Lambda/m_Q$). The strategy is
first to organise the sum over $n$ into an expansion in $v$ and then to
systematically compute $d \hat{\sigma}$ order-by-order in
$\alpha_s(m_Q)$. Technically, we do not a priori know where our efforts
are best placed, i.e. do we work at lowest order in $v$ and to NLO in
$\alpha_s$ or do we attempt to work at higher orders in $v$, but
computing each hard subprocess to lowest order? We need to know $v$ in
order to judge better what to do.

One final word before moving on to discuss $J/\psi$ photoproduction. 
For small $p_T$, NRQCD factorization is likely to break down, due to
contamination from higher twist effects. Also, one expects breakdown of
the NRQCD approach in the vicinity of the elastic scattering region,
i.e. $z \to 1$ where $z$ is the fraction of the photon energy carried by the
quarkonium (see later). 

\begin{figure}[!h]
\centerline{\epsfig{file=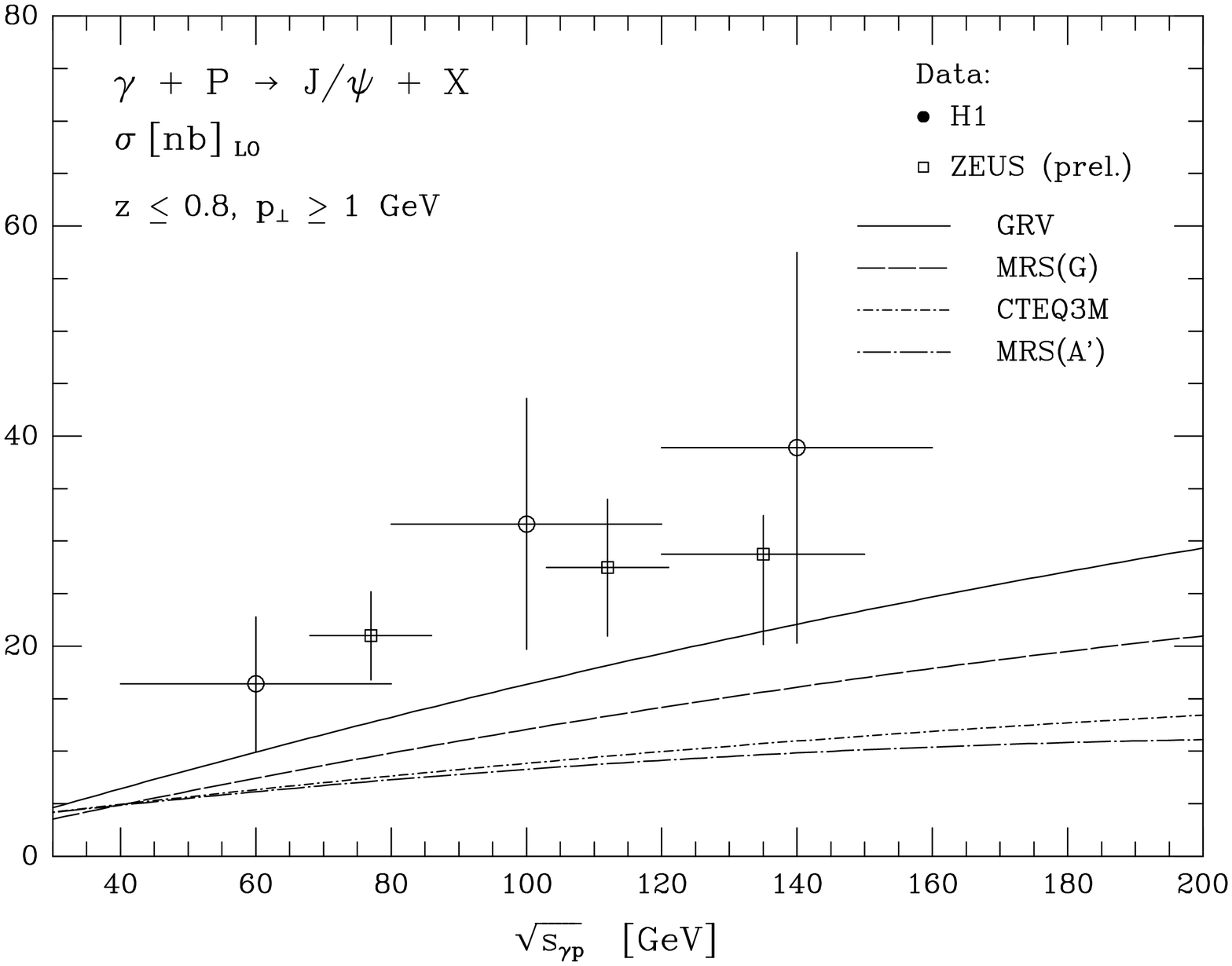,height=2.6in,width=2.4in,bbllx=26pt,%
bblly=89pt,bburx=551pt,bbury=764pt,angle=-90}}
\centerline{\epsfig{file=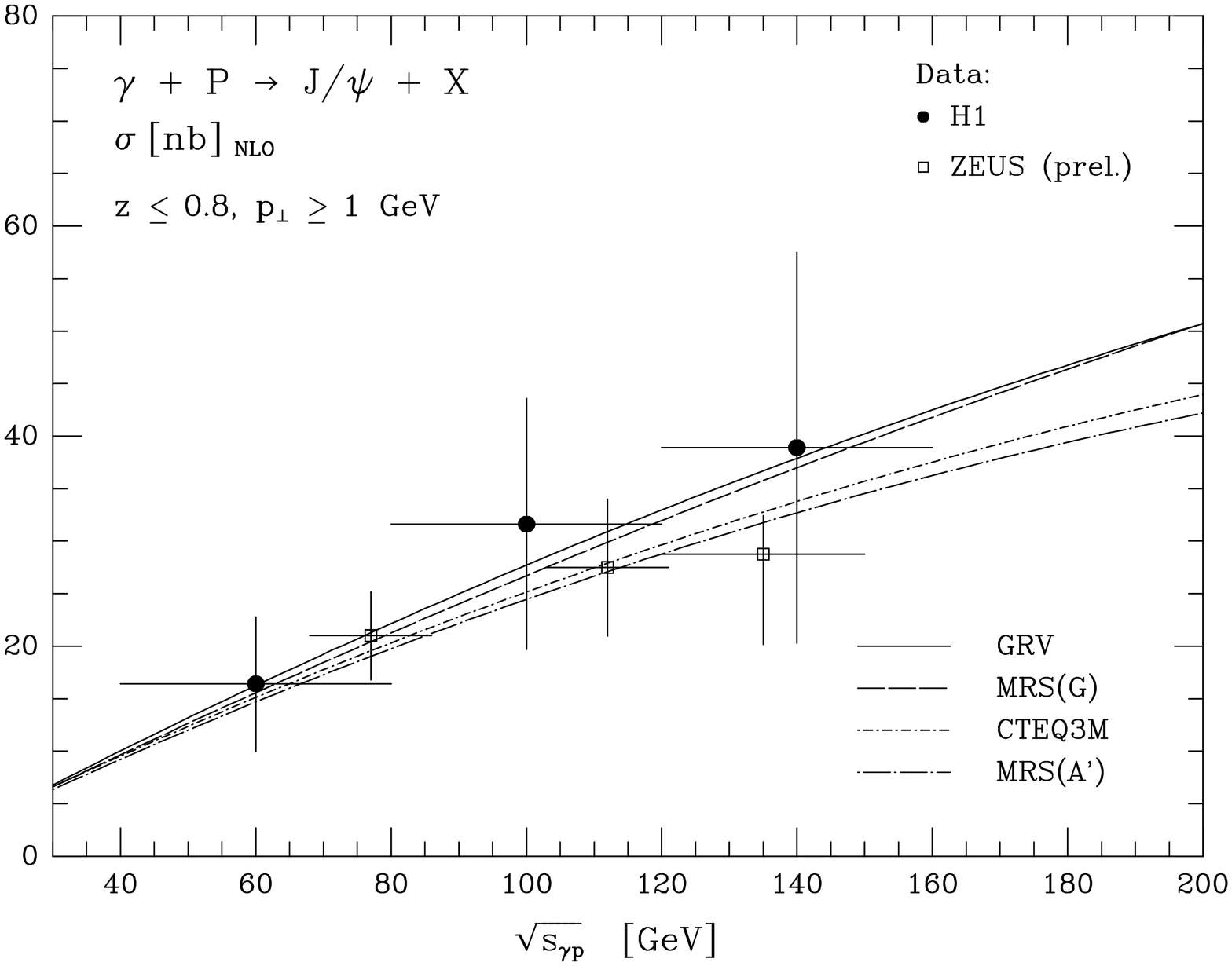,height=2.6in,width=2.4in,bbllx=26pt,%
bblly=89pt,bburx=551pt,bbury=764pt,angle=-90}}
\vspace{10pt}
\caption{LO and NLO colour-singlet predictions \protect\cite{MikeKramer} 
for the total inelastic
$J/\psi$ cross-section for different parameterizations of the proton
parton distribution functions compared to the HERA data \protect\cite{JPH1,JPZEUS}}
\label{fig9}
\end{figure}

Inelastic photoproduction of $J/\psi$ is something which has already
been measured at HERA. Let's see how the theory shapes up. To lowest
order in the velocity expansion, $[n] = [1,^3 \! S_1]$. The first entry in
the square brackets tells us that the $c \bar c$ is in a colour singlet
state, whilst the second entry tells us the spin and angular momentum of
the state. Not surprisingly, to lowest order in the velocity expansion,
the $c \bar{c}$ must be produced with the same quantum numbers as the
$J/\psi$. This is just the colour singlet model (CSM) of old. The lowest
order diagram which can contribute is shown in
Fig.\ref{fig8} and
$$ \langle O^{J/\psi}[1,^3 \! S_1] \rangle \sim | \phi(0) |^2 $$
where $\phi(0)$ is the wavefunction at the origin (it can be extracted
from the electronic width of the $J/\psi$). NLO$(\alpha_s)$ corrections
have been computed \cite{ZSZK} and shown to be large. Fig.\ref{fig9} shows that
the NLO corrections enhance the LO calculation and lead to a reduced 
sensitivity to the gluon density in the proton. 

\begin{figure}[!h] 
\centerline{\epsfig{file=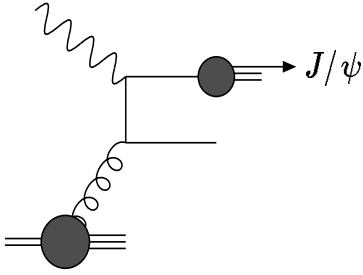,height=1.5in,width=2.0in,%
bbllx=169pt,bblly=388pt,bburx=296pt,bbury=475pt}}
\vspace{10pt}
\caption{Fragmentation contribution to the production of the $J/\psi$.}
\label{fig10}
\end{figure}

One might well ask as to the significance of the resolved photon
contribution. It is important at small enough $z$ \cite{GRS,CnK,KnK}.
In addition, for mesons produced at high enough $p_T$ we have an
additional scale to consider and terms which are leading in $\alpha_s$
can be suppressed by powers of $\sim m_c^2/p_T^2$. This is
true for example of the diagram shown in Fig.\ref{fig8} relative
to that shown in Fig.\ref{fig10}. 
The latter fragmentation contribution is higher order in
$\alpha_s$, however there is one less hard quark propagator and so it
will dominate for large enough $p_T$. Fragmentation contributions and
resolved photon contributions are not important in computing the total
rate for $z > 0.4$ (which is essentially where the data is), and so we
expect the curves in Fig.\ref{fig8} to be reliable. 

\begin{figure}[!h] 
\centerline{\epsfig{file=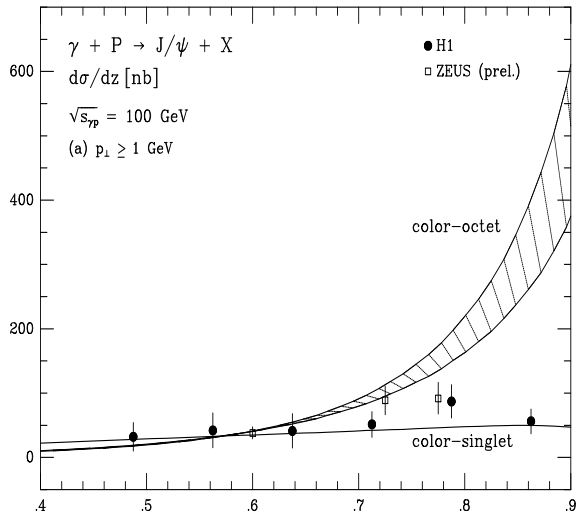,height=3.0in,width=2.7in,bbllx=30pt,%
bblly=78pt,bburx=519pt,bbury=755pt,angle=-90}}
\vspace{10pt}
\caption{Colour singlet and colour octet contributions to the $J/\psi$
energy distribution \protect\cite{MikeKramer} compared with the HERA data 
\protect\cite{JPH1,JPZEUS}.}
\label{fig11}
\end{figure}

Going to NLO in the velocity expansion means moving away from the
CSM. For the first time we encounter colour octet contributions. In
particular, the LO$(\alpha_s)$ diagram is again that of
Fig.\ref{fig8}\footnote{There is a lower order contribution in which 
no gluon is
radiated off, however this would give a contribution only at $z=1$ which
lies outside our region of interest.}
now the $c \bar{c}$ can be formed in one of 5 states, i.e.
$[n] = [8,^1 \! S_0], [8,^3 \! S_1], [8,^3 \! P_{0,1,2}]$. 
The price one pays for
having to convert this state into the $J/\psi$ is an extra power of
$v^4$ relative to the colour singlet matrix element, i.e.
$$ \langle O^{J/\psi}[n] \rangle  \sim v^4 \langle O^{J/\psi}[1,^3 \! S_1]
\rangle.$$ It turns out that the $v^4 \sim 0.01$ suppression of the long
distance matrix elements is partially compensated for by a strong
enhancement of
the corresponding short distance cross-section. In particular, this is
so for the $[8,^1 \! S_0]$ and $[8,^3 \! P_{0,2}]$ states. The colour singlet
matrix element can be extracted from data, e.g. the leptonic width of the
$J/\psi$. Likewise, we need to fit these new matrix elements to data (or
extract them from lattice calculations). It
is therefore clear that a test of the NRQD framework requires
data from different sources -- the challenge being to find a consistent
description. This is a particularly topical issue, since an explanation
of the Tevatron excess of direct $J/\psi$ and $\psi'$ production needs, in
addition to fragmentation contributions, colour octet contributions
\cite{BF}. One
can use the Tevatron data to fit the relevant matrix elements. The validity
of this explanation can be confirmed (or not) on comparing to data
which can be obtained from HERA. Unfortunately, the matrix elements
which are important at the Tevatron are not so important in $J/\psi$
photoproduction for $z > 0.4$.  
However, the key matrix element for the Tevatron ($[8,^3S_1]$) does play
a key role in the region of lower $z$, where (large enough $p_T$) the
dominant contribution comes from the fragmentation mechanism via
resolved photons \cite{CnK,KnK}.    
Another process which is sensitive to the $[8,^3S_1]$ state is
the photoproduction of $J/\psi + \gamma$ (where the photon is produced
in the hard subprocess, i.e. not via the radiative decay of a $P$-wave
quarkonium) \cite{CGK}. So, with the anticipated increase in statistics, we
can really expect to test NRQCD at HERA. 
Going back to the $J/\psi$, there are some weak constraints
on the important matrix elements from the Tevatron data and these have been
used in the theoretical calculations of \cite{Mike}. In Fig.\ref{fig11} 
the HERA
data on the $z$ distribution are shown and seen to compare very well
with the colour singlet calculation. The colour octet contribution
however, is much too large at large $z$. The shaded band on the theory
prediction denotes the type of uncertainty expected from the fit to the matrix
elements, although this error is hard to ascertain with confidence.  
Thus the HERA data is not supporting a large colour octet
contribution at large $z$. However, one must be careful in interpreting this as
evidence against the NRQCD approach, since the $z \to 1$ region is
sensitive to higher order non-perturbative contributions which lead to
the breakdown of the NRQCD expansion \cite{BRW}.

In addition to the processes just discussed, increased statistics will
allow measurement of other meson states, e.g. $\psi'$ and $\Upsilon$
which will certainly further test our understanding of QCD. 

Before finishing, I'd like to mention an alternative to the NRQCD
approach which is the colour evaporation model, e.g. see \cite{Halzen} 
(essentially a resurrection of the old duality calculations \cite{dual}). 
The production rate is calculated according to
\begin{equation}
\sigma = \frac{1}{9} \int_{2m_c}^{2m_D} dm \frac{d \sigma(c
\bar{c})}{dm},
\end{equation}
where $m$ is the invariant mass of the produced $c \bar{c}$ pair.
This is the cross-section for production of charmonium, i.e. it is the
sum of production rates for $J/\psi, \psi', \chi$ and $\eta_c$.  
To get the rate for any particular state, one must multiply by some
(empirically determined) factor. The factor
of $1/9$ comes from arguing that the colour octet $c \bar{c}$ convert
with probability 1/9 into a singlet state (which must be charmonium in
this mass range). The colour bleaching is supposed to
occur as a result of some soft final state interactions, the precise
nature of which is not important. For more details on the
phenomenological success of the colour evaporation approach, I refer to
\cite{Halzen} and references therein.

\section{For the future...}
The future bodes well, not least as a consequence of the anticipated
increase in statistics. Higher statistics will mean more precise studies
of open charm production over a wide kinematic range (which is vital if
we are to develop our understanding) and will also allow us to really test
our understanding of charmonium production and, in particular, NRQCD.
The intriguing issue of forward jet production needs further work before
we can claim to really understand what is happening. NLO theory
facilitates more precise tests of single and dijet rates using what
is becoming very good data -- it is true to say that we are learning
how to really test the theory. 
I didn't mention the structure of the virtual photon or prompt photon
production. Good data is starting to accumulate on both these processes
and we can anticipate that much will be learnt over the next few years.

\section*{Acknowledgements}
I sincerely thank Jon Butterworth, Michael Kr\"amer and Mike Seymour for
their help in preparing this talk....it was invaluable. Thanks also to
Matteo Cacciari and Michael Klasen for providing me with Fig.7 and
Figs.2 \& 3 respectively.

\end{document}